\documentclass[lettersize,journal]{IEEEtran}
\usepackage{amsmath,amsfonts}
\usepackage{algorithmic}
\usepackage{algorithm}
\usepackage{array}
\usepackage{booktabs}
\usepackage[caption=false,font=footnotesize,labelfont=rm,textfont=rm]{subfig}
\usepackage{textcomp}
\usepackage{stfloats}
\usepackage{url}
\usepackage{verbatim}
\usepackage{graphicx}
\usepackage{float} 
\usepackage{cite}
\usepackage{amsthm}
\usepackage{amssymb}
\usepackage{xcolor}
\usepackage{dsfont}
\usepackage{soul}
\usepackage{makecell}
\usepackage{bbding}
\newlength \figwidth
\definecolor{bittersweet}{rgb}{1.0, 0.44, 0.37}
\definecolor{glaucous}{rgb}{0.38, 0.51, 0.71}
\definecolor{gainsboro}{rgb}{0.86, 0.86, 0.86}
\definecolor{babyblueeyes}{rgb}{0.63, 0.79, 0.95}
\definecolor{silver}{rgb}{0.75, 0.75, 0.75}
\definecolor{neoncarrot}{rgb}{1.0, 0.64, 0.26}
\definecolor{Gray}{gray}{0.6}
\definecolor{LightCyan}{rgb}{0.88,1,1}
\definecolor{BackgroundLightBlue}{rgb}{0.97,0.97,1}
\definecolor{BackgroundGray}{gray}{0.98}

\def\nb0{{\mathbf{0}}}
\def\nb1{{\mathbf{1}}}

\begin{document}


\title{Towards Ubiquitous 6G Computing and Networking Convergence: Architecture and Mechanism for Cross-Domain Resource Coordination}

\author{
Yang Li,
Xing Zhang,
Yan Zhang,
and Wenbo Wang

\thanks{Yang Li, Xing Zhang (Corresponding author), and Wenbo Wang are with the School of Information and Communications Engineering, Beijing University of Posts and Telecommunications, China.}
\thanks{Yan Zhang is with the School of Information and Communication Engineering, University of Electronic Science and Technology of China.}
}



\maketitle

\begin{abstract}
The 6G network will support six major application scenarios, such as immersive communication, integrated AI and communication, and integrated sensing and communication. Many scenarios necessitate significant computational support. Moreover, user demands are becoming increasingly segmented, diverse, and personalized. Traditional network slicing alone is insufficient to meet the heterogeneous computing and networking demands of emerging service scenarios. Mobile computing-network convergence (CNC) introduces a fundamentally different paradigm from the conventional “cloud computing plus communication network” model by deeply embedding computing resources into the mobile network infrastructure and enabling integrated computing–network services tailored to diverse user demands. In this article, we investigate orchestration architectures and mechanisms for CNC in 6G mobile networks. We begin by reviewing the evolution of CNC from a mobile network perspective and surveying existing studies, which we categorize according to mobile network architectures. Building on these insights, we propose a hierarchical, cross-domain coordination architecture and an orchestration mechanism based on hierarchical multi-agent reinforcement learning. Performance evaluations demonstrate that the proposed architecture and mechanism significantly reduce system energy consumption while enhancing task satisfaction rate. Finally, we discuss open challenges and future research directions.

\end{abstract}


\section{Introduction}

As mobile networks evolve from 1G voice-only services to 5G supporting eMBB, mMTC, and URLLC \cite{1}, and further toward the six emblematic 6G application scenarios defined by ITU \cite{2}, it becomes evident that service paradigms are continuously transforming, driven by increasingly diverse user demands. With the advent of emerging application scenarios such as immersive interactions, vehicle-to-everything (V2X), and smart industry, mobile communications are gradually transcending the traditional boundaries of bandwidth and connectivity. This shift is propelling networks to evolve from a purely connection-oriented paradigm to a more holistic service-oriented architecture. Against this backdrop, the deep convergence of computing and networking has emerged as a pivotal enabler for future intelligent service delivery and sustainable network evolution.

\begin{figure*}[!t]
\centering
\includegraphics[width=1.8\figwidth]{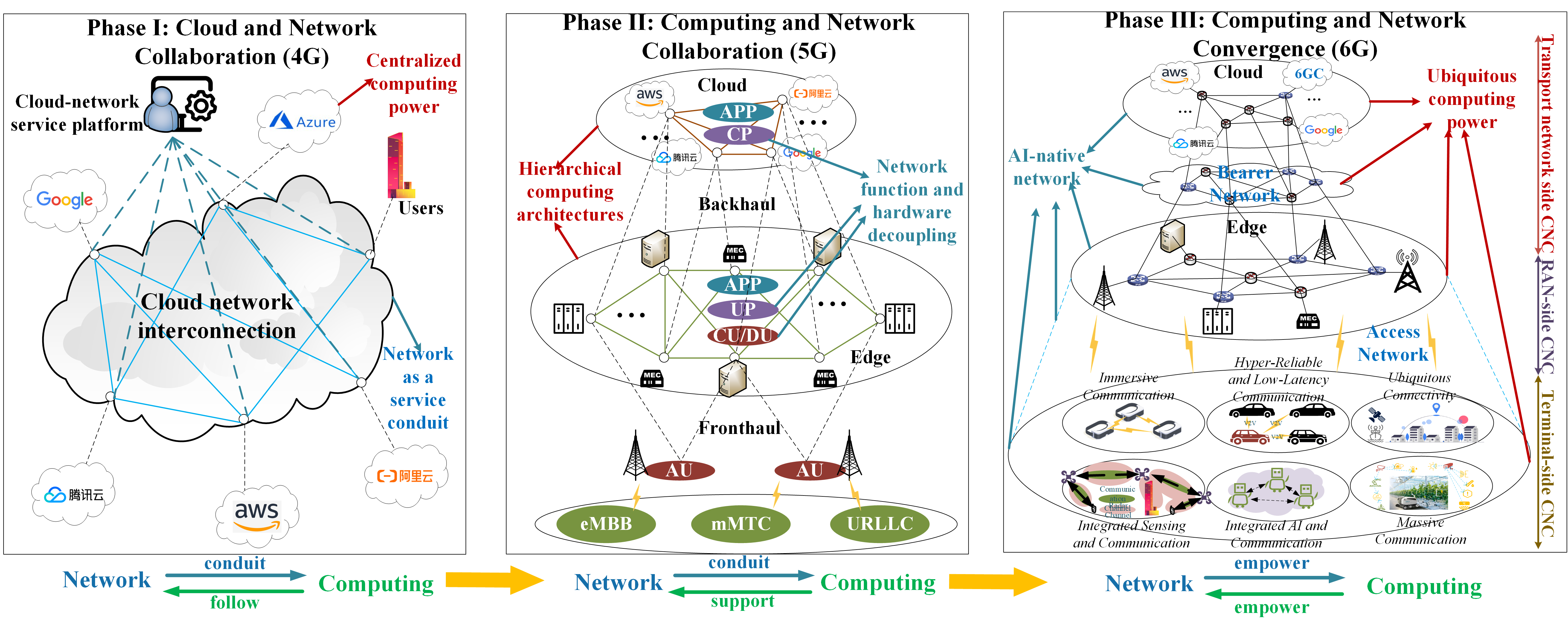}
\caption{The three evolutionary stages of computing and network convergence.}
\label{fig:1}
\vspace{-0.5cm}
\end{figure*}

The computing and networking convergence (CNC) can be categorized into three stages, as illustrated in Fig. \ref{fig:1}. The first stage, the era of cloud–network collaboration, began in 3G and reached large-scale deployment in 4G. During this phase, computing and networking remained relatively independent systems with separate orchestration and scheduling \cite{3}, yet began to evolve toward collaborative deployment and operation. Networks primarily provide connectivity for computing services, while computing resource provisioning is constrained by network bottlenecks and dynamic task demands.

The second stage, the era of computing–network collaboration, emerged with 5G. Driven by the continuous advancement of key technologies such as network function virtualization (NFV), software-defined networking (SDN), and cloud-native architectures, network functions that were traditionally bound to dedicated hardware have been progressively decoupled from the underlying infrastructure. These network functions can now operate as containerized or virtual machine instances on general-purpose computing platforms. This evolution has driven the transformation of networks from static, closed architectures to flexible and programmable frameworks, establishing the foundation for dynamic deployment and intelligent orchestration of network functions. Meanwhile, edge and cloud computing have formed a layered and collaborative ecosystem that jointly enables service-oriented distributed computing, addressing diverse requirements for low latency and high energy efficiency. At this stage, the network primarily offers connectivity capabilities, while computing resources start to reinforce the network infrastructure, enabling elastic scaling and on-demand migration of network functions. The collaboration between computing and networking has deepened, marking a significant step toward their functional convergence.

The third stage represents an era where computing–network integration becomes a native capability of the network, and is anticipated to be one of the defining characteristics of 6G networks. At this stage, a hierarchical computing architecture spanning the cloud, edge, and device layers will enable ubiquitous computing capabilities. Computing resources will not only sustain the efficient operation and agile evolution of network functions but also drive the network’s intelligent transformation through deep integration and mutual empowerment. The 6G network will exhibit AI-native characteristics, enhancing its capability to optimize computing resource management through intelligent perception, analysis, and decision-making. Within this bidirectionally enhanced closed-loop system, computing and networking will transcend simple collaboration to realize intrinsic convergence, forming a unified service-oriented resource pool.

However, 6G CNC still faces multiple critical challenges in practical implementation. First, user service demands exhibit extreme diversity and pronounced spatio-temporal heterogeneity, leading to highly dynamic uncertainties in supply-demand matching. Second, the underlying infrastructure is evolving toward a heterogeneous and distributed deployment paradigm. The marked heterogeneity of computing architectures, network capabilities, and operational modes across cloud, edge, and device layers, along with the dynamic nature of their states, makes unified modeling and efficient global management particularly challenging. Furthermore, on the service supply side, following the Service-Based Architecture (SBA) in 5G, 6G CNC is expected to fully adopt a cloud-native microservices paradigm. In this context, fine-grained service instances and their increasingly intricate interdependencies further exacerbate the challenges of service orchestration.

To address these challenges, this paper proposes a hierarchical, cross-domain coordination architecture and an intelligent orchestration mechanism for 6G CNC. The goal is to achieve efficient coordination among the user side, the infrastructure side, and the service provisioning side through dynamic sensing, intelligent scheduling, and adaptive orchestration of computing, networking, and service resources. The main contributions of this paper are summarized as follows:
\begin{itemize}
    \item A comprehensive survey of existing research efforts on CNC is conducted, revealing current research gaps.

    \item A hierarchical, cross-domain coordination architecture and an orchestration mechanism based on hierarchical multi-agent reinforcement learning (HMARL) is proposed. Its effectiveness is validated through extensive performance evaluations.

    \item Potential challenges and future research directions are identified and discussed to provide guidance for the ongoing evolution and practical deployment of 6G CNC.
\end{itemize}


\begin{table*}[t]
\caption{Summary of current research progress on CNC}
\label{tab:1}
\resizebox{\textwidth}{!}{%
\begin{tabular}{l|l|l|l}
\toprule[1.5pt]
Aspect                             & Terminal-Side CNC \cite{5,6,7}                                                                                                                 & RAN-Side CNC \cite{8,9,10}                                                                                                                                   & Transport Network Side CNC \cite{11,12,13}                                                                                                           \\ \toprule[1.5pt]
Constituent Elements               & \begin{tabular}[c]{@{}l@{}}User devices and device-to-device \\ networks\end{tabular}                                          & \begin{tabular}[c]{@{}l@{}}Wireless access points (e.g., base \\ stations, roadside units), access \\networks, and  edge computing nodes\end{tabular} & \begin{tabular}[c]{@{}l@{}}Backbone networks, transport \\ networks, core networks, \\ and cloud-edge infrastructures\end{tabular}      \\ \hline
Computing Resource Characteristics & \begin{tabular}[c]{@{}l@{}}Highly fragmented, low-power, \\ and densely distributed\end{tabular}                               & \begin{tabular}[c]{@{}l@{}}Moderate computing capability with \\ distributed characteristics\end{tabular}                                 & \begin{tabular}[c]{@{}l@{}}High-performance centralized \\ computing capability (e.g., \\ edge clouds, core clouds)\end{tabular}        \\ \hline
Network Resource Characteristics   & \begin{tabular}[c]{@{}l@{}}Short-range, dynamic connections, \\ and multi-protocol access support\end{tabular}                 & \begin{tabular}[c]{@{}l@{}}Heterogeneous wireless access and\\ wide coverage\end{tabular}                                   & \begin{tabular}[c]{@{}l@{}}High-bandwidth and relatively \\ stable network topology\end{tabular}                                      \\ \hline
Main Application Scenarios         & \begin{tabular}[c]{@{}l@{}}On-device AI inference, \\ privacy-preserving computation, \\ and device collaboration\end{tabular} & \begin{tabular}[c]{@{}l@{}}Video pre-processing, edge caching, \\ and latency-sensitive services\end{tabular}                                     & \begin{tabular}[c]{@{}l@{}}
Large-scale AI model training, \\digital-twin based simulation, and \\ decentralized storage\end{tabular}               \\ \hline
Challenges and Limitations         & \begin{tabular}[c]{@{}l@{}}Constrained energy resources, \\ high heterogeneity, \\ and dynamic network topology\end{tabular}   & \begin{tabular}[c]{@{}l@{}}Diverse access modes, fluctuating \\ wireless resources, and limited \\ computing capability\end{tabular}                            & \begin{tabular}[c]{@{}l@{}}High bandwidth pressure and \\ complexity of unified \\ coordination and scheduling\end{tabular}             \\ \toprule[1.5pt]
\end{tabular}%
}
\end{table*}

\section{Research Progress on CNC}

Recently, multiple studies on CNC have emerged \cite{5,6,7,8,9,10,11,12,13,14,15}. These works are summarized and categorized in Table \ref{tab:1}.

\subsubsection{Terminal-Side CNC}
With the continuous enhancement of terminal capabilities, the increasing amount of ubiquitous and distributed terminal-side computing power remains underutilized. Consequently, several studies investigate the architecture, key technologies, and application scenarios of terminal-side CNC \cite{5,6,7}. In \cite{5}, a multi-granularity, multi-level strategy for computing power scheduling and microservice deployment was proposed for end-side computing power networks. In \cite{6}, the authors developed a terminal-side computing force network architecture for large-scale heterogeneous terminals, along with an Android-based cloud-native container resource scheduling scheme. In \cite{7}, the architecture and key technologies of terminal-side CNC were analyzed, and simulation results demonstrated its significant potential.

Although terminal-side CNC shows great potential for improving the utilization efficiency of distributed terminal-side computing resources, several critical challenges remain. First, terminals from different vendors tend to form isolated ecosystems, resulting in poor cross-brand collaboration capabilities. Second, coordinating computing and networking across massive numbers of heterogeneous terminals requires lightweight protocols that can operate effectively in resource-constrained device environments. Finally, privacy preservation and security assurance remain core challenges requiring urgent attention in terminal-side collaboration.

\subsubsection{RAN-Side CNC}
Driven by emerging application scenarios, radio access networks (RANs) have adopted edge and fog computing technologies while continuously evolving toward cloud-based, open, and intelligent architectures. This evolution has gradually enhanced the computational capabilities of access networks. Against this backdrop, the deep integration and on-demand provisioning of communication and computing resources have emerged as key technological trends in the evolution of RANs.

References \cite{8,9,10} investigate the architectures and key technologies for RAN-side CNC. In \cite{8}, the authors proposed an ML-based resource allocation framework under the O-RAN architecture to enable joint optimization of communication and computing resources. In \cite{9}, a DRL-based RAN slicing mechanism was developed for heterogeneous service demands. In \cite{10}, a new paradigm of CNC in RANs was explored. The study systematically outlined its design principles, logical architecture, and key technologies.

RAN-side CNC primarily focuses on coordination among the intrinsic computing capabilities of access nodes, externally attached computing resources (e.g., edge computing nodes), and access network resources. Although computational resources in access networks show a certain degree of centralization, they remain limited overall and exhibit heterogeneous and distributed characteristics. Mobility presents a major challenge for RAN-side CNC. To ensure continuous computing and communication services within the RAN, user handover mechanisms and service migration strategies need to be jointly designed. Furthermore, the RAN's capability to support deterministic services must be strengthened, since the wireless access layer is highly susceptible to environmental variations, making transmission quality difficult to guarantee.

\subsubsection{Transport Network Side CNC}
In 6G, computing-centric services are expected to become ubiquitous across the network. As a critical bridge enabling collaborative computing between cloud data centers and edge computing nodes, the transport network plays a pivotal role in supporting end-to-end (E2E) computing–network integration. Consequently, transport network side CNC technologies exemplified by computing power networks (CPNs) have attracted growing attention \cite{11,12,13}.

Reference \cite{11} provided a comprehensive survey of CPN, highlighting the integration of ubiquitous and heterogeneous computing resources through network interconnection. Reference \cite{12} analyzed the convergence of network–cloud/edge computing, focusing on key technologies, challenges, architectural models, and standardization progress. In \cite{13}, the architecture of CPNs was investigated, with an emphasis on key technologies and application scenarios.

Although transport network side CNC benefits from a stable network topology and relatively centralized computing resources, it still faces several challenges, including heavy traffic loads and substantial overhead caused by network-wide synchronization of computing resource information.

As summarized above, existing studies mainly concentrate on terminal-side CNC, RAN-side CNC, and transport-network-side CNC, each within its respective architectural scope. However, 6G inherently calls for E2E CNC service capabilities to support diverse and dynamic user requirements. Most existing work lacks a systematic investigation of cross-layer and cross-domain collaborative mechanisms, hindering the global realization of CNC and leading to fragmented resource utilization and suboptimal E2E service performance.

Although existing end–edge–cloud collaborative computing approaches improve resource utilization to some extent \cite{14,15}, their design paradigms exhibit key limitations. First, the network is often treated as a passive data transmission pipeline, lacking joint optimization with computing resources. Second, most approaches focus on localized coordination, making it difficult to guarantee E2E service performance. In contrast, this paper presents a CNC framework that enables deep co-optimization of computing and network resources. The network is transformed from a traditional transport pipe into an orchestratable resource participating in service deployment and task scheduling. The framework coordinates the E2E orchestrator and domain orchestrators to achieve global optimization across terminal, RAN, and transport domains. In addition, an HMARL mechanism integrated into the framework continuously adapts policies in dynamic environments, improving system adaptability and overall performance.

\section{Architecture and Mechanism for Cross-Domain Resource Coordination}

\begin{figure}[!t]
\centering
\includegraphics[width=\figwidth]{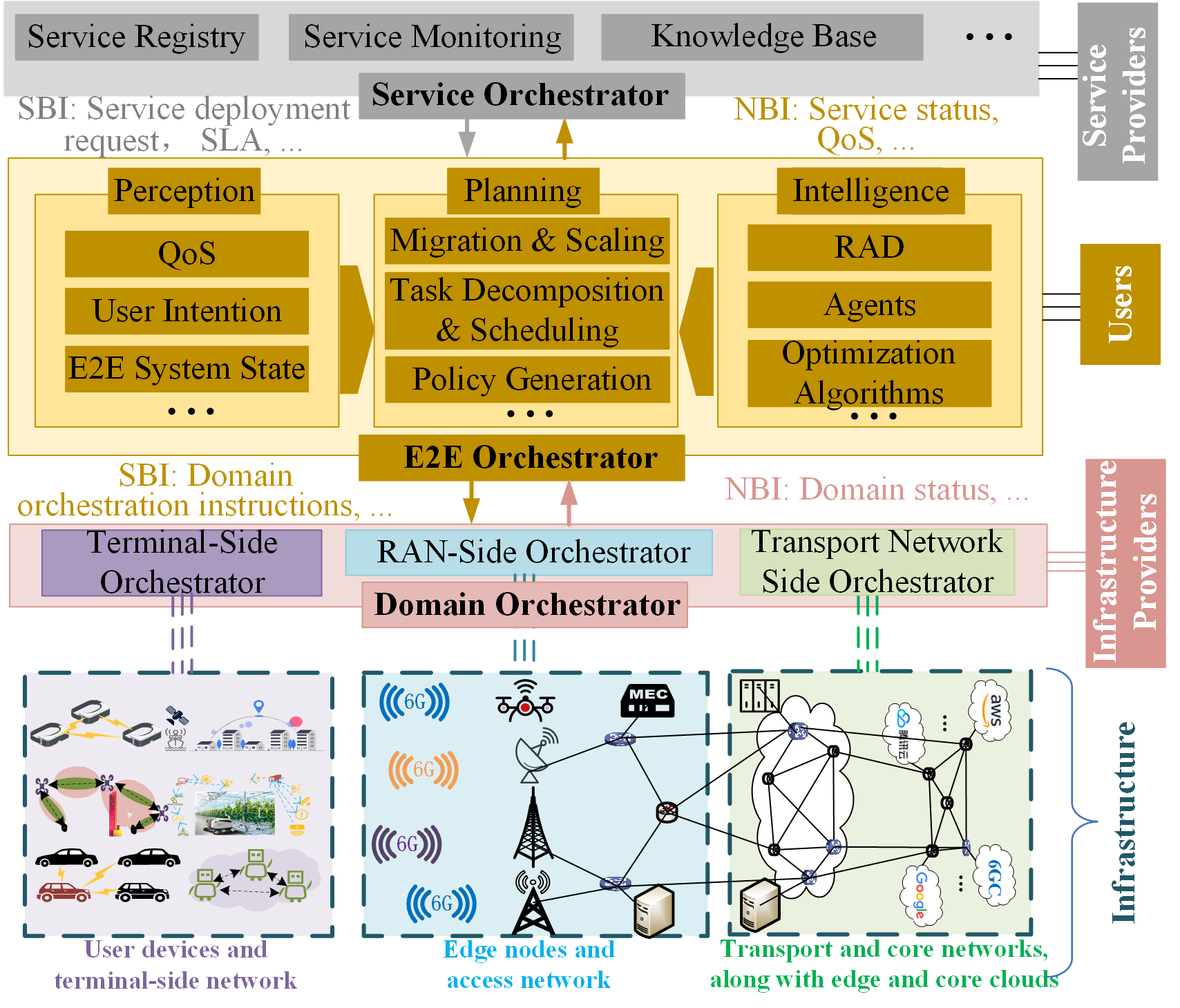}
\caption{The Architecture for cross-domain CNC orchestration.}
\label{Fig: 2}
\vspace{-0.5cm}
\end{figure}

The 6G ecosystem consists of three main entities: infrastructure providers, service providers, and users. Infrastructure providers deliver the physical resources necessary to support network functions and service applications. Users may request one or more services. Each request is scheduled for execution in a container or virtual machine hosting the corresponding service. Service providers develop and maintain various services, which are deployed on physical infrastructure via containers or virtual machines. Achieving 6G CNC requires effective coordination among the three parties to ensure efficient matching of physical resources, service instances, and user requests. Accordingly, we propose a hierarchical cross-domain CNC orchestration framework, as illustrated in Fig. \ref{Fig: 2}, and develop an HMARL-based orchestration mechanism upon it. The proposed architecture and mechanism are built upon two core principles: (i) enabling hierarchical cross-domain collaboration and intra-domain autonomy to achieve coordinated orchestration of heterogeneous computing and networking resources; and (ii) leveraging online learning to continuously optimize decision-making policies, thereby allowing the system to proactively adapt to dynamic environmental variations. 

\vspace{-0.15cm}
\subsection{Service Orchestrator}
\vspace{-0.05cm}
The service orchestrator, associated with service providers, manages the full lifecycle of services. It receives service information from providers and issues decision directives to the E2E orchestrator for execution via the southbound interface (SBI). Specifically, it performs the following core functions:

\subsubsection{Service Registration}
This module receives, parses, and persistently stores registration requests from service providers, enabling service onboarding and lifecycle management. Registration requests include essential metadata such as service instances, workflow graphs, data models, service-level agreement (SLA) requirements, and compute/storage demands.

\subsubsection{Service Monitoring}
This module continuously monitors all deployed services from an E2E perspective, maintains real-time QoS abstractions, and feeds relevant state information back to service providers.

\subsubsection{Knowledge Base}
This module provides unified management of heterogeneous knowledge and operational information, such as historical runtime data, pre-trained models, and expert knowledge. In the absence of sufficient real-time data for online training, it supports orchestration decisions by leveraging pre-trained models and expert knowledge.

\begin{figure*}[!t]
\centering
\includegraphics[width=1.6\figwidth]{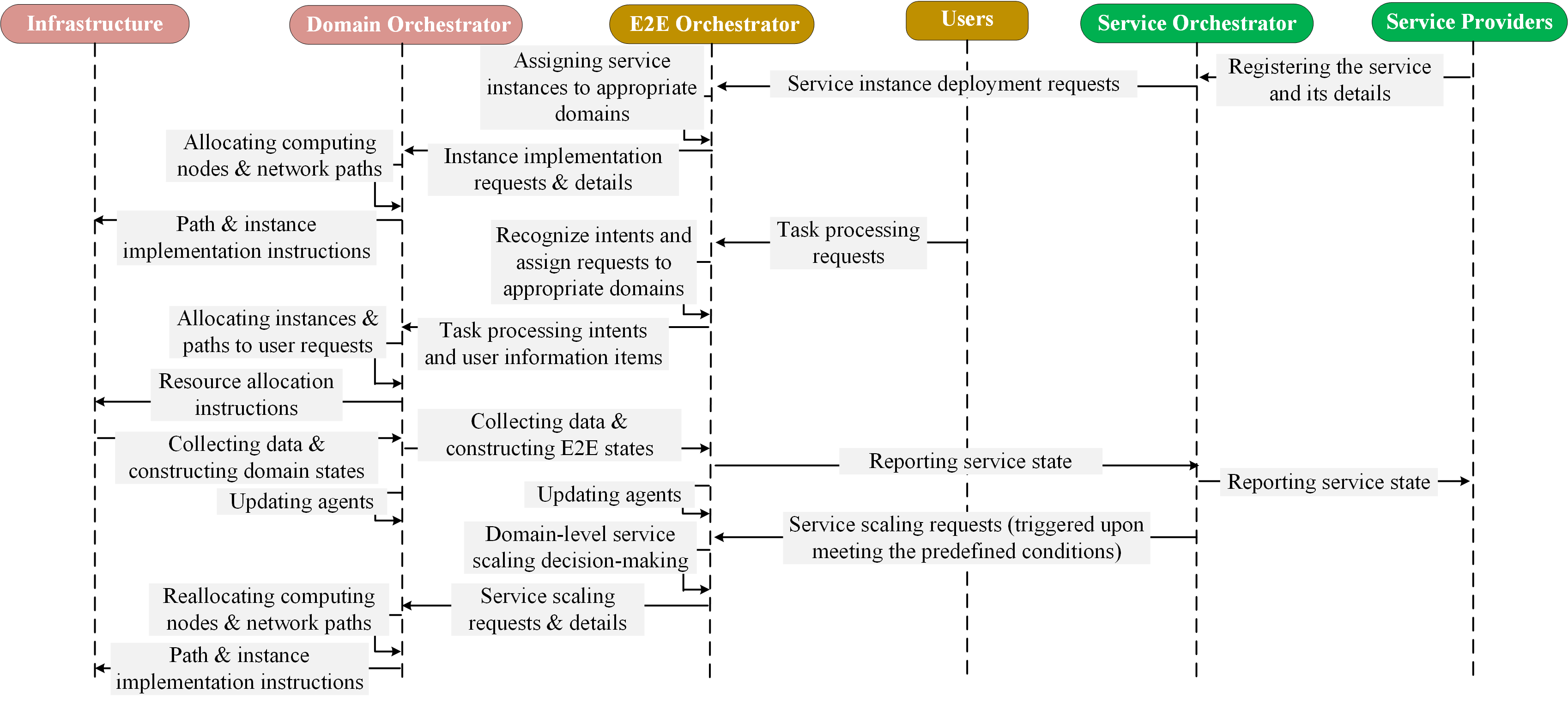}
\caption{The flowchart of E2E orchestration.}
\label{Fig:3}
\vspace{-0.5cm}
\end{figure*}

\subsection{E2E Orchestrator}

The E2E orchestration layer is associated with users and serves as a critical component for enabling cross-domain coordination. It exposes capabilities to the service orchestration layer via the northbound interface (NBI) and invokes functionalities from domain orchestration layers through the SBI. Its primary responsibilities include the following:

\subsubsection{Perception and Intent Recognition}
To meet diverse 6G service demands and enable accurate decision-making, CNC systems require comprehensive perception of both network states and user intent. On one hand, critical information such as distributed computing resources, network topology, and link status must be aggregated, forming the basis for intelligent decision-making. On the other hand, user intent, including behavioral patterns, QoS requirements, and service priorities, should be integrated with system status to support task scheduling and service orchestration. 

\subsubsection{Cross-domain Coordination and Planning}
As a core function of the E2E orchestration layer, this module coordinates domain orchestrators across the terminal-side, RAN-side, and transport network side in the horizontal dimension. Built upon the cross-domain state awareness enabled by the perception module, this component facilitates holistic planning, resource allocation, and workflow orchestration for E2E CNC services. For cross-domain orchestration requests from the service orchestration layer, the E2E orchestration layer formulates optimal deployment and routing strategies based on global resource availability and network topology. It must also rapidly respond to customized user demands by leveraging multi-domain resources to construct E2E service chains that satisfy latency, bandwidth, and computing constraints, ensuring both service quality and system efficiency.

\subsubsection{Intelligent Driving and Optimization}
This module provides knowledge-enhanced adaptive decision-making for E2E service orchestration and task scheduling. Specifically, a retrieval-augmented decision-making (RAD) mechanism extracts expert knowledge from the knowledge base to assist agents in making informed orchestration decisions. Meanwhile, a DRL mechanism continuously updates policies based on runtime feedback, enabling adaptive optimization in dynamic environments. By integrating knowledge-driven and data-driven approaches, agents balance decision stability and environmental adaptability, thereby improving long-term service performance and resource utilization efficiency.

\vspace{-0.2cm}
\subsection{Domain Orchestrator}
The domain orchestration layer is associated with infrastructure providers and manages computational and network resources within each domain. Through its NBI, it reports status to the E2E orchestration layer, and via its SBI, it invokes capability interfaces exposed by infrastructure providers. Due to the pronounced heterogeneity across domains in terms of architectures, interface standards, and data models, the domain orchestration layer is required to perform unified semantic modeling of underlying resources and map them into standardized representations interpretable by the E2E orchestration layer, thereby enabling composable and schedulable cross-domain resources. Specifically, the three types of domain orchestrators play distinct yet complementary roles in cross-domain coordination.

\subsubsection{Terminal-Side Orchestrator}
Manages terminal devices and local resources, supporting lightweight task execution and on-device inference, with an emphasis on low latency and efficient operation under resource-constrained environments.

\subsubsection{RAN-Side Orchestrator}
Coordinates radio access and edge computing resources to enable joint communication–computing scheduling, accommodating mobility and time-varying network conditions.

\subsubsection{Transport Network Side Orchestrator}
Provides large-scale computing capacity and a global network view based on core network and cloud infrastructure, supporting cross-region resource scheduling and high-throughput task processing.

\subsection{Workflow}

As illustrated in Fig. \ref{Fig:3}, the service provisioning process in the proposed CNC architecture starts with the service provider registering the service and its specific characteristics within the service orchestration layer. Subsequently, service providers can submit service instance deployment requests, which are forwarded to the E2E orchestrator. Upon identifying the service characteristics and corresponding SLA requirements, the E2E orchestration layer delegates the service instance to the appropriate domains according to the real-time domain states. The domain orchestrators then manage domain-specific resource allocation tasks, such as selecting compute nodes and planning network paths. Finally, deployment instructions and path configurations are issued to the underlying computing and networking infrastructure, thereby completing the service deployment process.

Users can dynamically initiate task processing requests associated with registered services. The E2E orchestrator first identifies the user intent and maps the request to the appropriate processing domains. Upon receiving the task processing request and user intent, domain orchestrators select service instances and network paths to ensure efficient task execution.

Meanwhile, domain orchestrators continuously sense and collect status information from intra-domain resources and services to construct domain-level states, which are then reported to the E2E orchestrator. The E2E orchestrator further aggregates multi-domain information to establish a global system view and provides the required service information to the service orchestration layer.

Once the service orchestrator detects that the service state meets predefined re-orchestration conditions, it notifies the E2E orchestrator. Typical triggers include persistent SLA violations in high-percentile E2E latency, sustained overload of critical CPU/GPU resources, and significant shifts in traffic hotspots. Based on the global system view, the E2E orchestrator generates re-orchestration decisions. These decisions are then dispatched to the relevant domain orchestrators, which reallocate compute nodes and network paths within their respective domains, thereby ensuring continuous alignment between resource allocation, service load, and the spatiotemporal distribution of user requests.

\subsection{Mechanism}
In the above process, the E2E orchestrator and domain orchestrators undertake global decision-making and local execution, respectively. This process can be naturally formulated as a hierarchical multi-agent reinforcement learning (HMARL) problem \cite{16}, where orchestration entities at different layers are modeled as collaborative agents to enable cross-domain resource coordination and online optimization.

Specifically, the E2E orchestrator is modeled as a high-level distributor agent that makes decisions such as cross-domain task allocation and service deployment based on the global system state, with the objective of maximizing system-wide performance metrics. In contrast, domain orchestrators are modeled as multiple low-level executor agents that, upon receiving high-level decisions, perform fine-grained actions based on local states such as compute node selection, network path planning, and task scheduling. These low-level agents operate independently within their domains while coordinating through high-level policies and feedback.

Under this HMARL framework, system operation is cast as a continuous interaction and learning process. A global reward is constructed based on system-wide performance metrics and further decomposed into local rewards to guide policy updates for both high-level and low-level agents. Through reward decomposition and collaborative learning, the framework achieves coordinated optimization across layers. Overall, this mechanism establishes a closed-loop, learning-driven orchestration framework that unifies perception, decision-making, execution, and feedback, enabling cross-domain coordination and long-term performance optimization.
\section{Case Study}

\begin{figure}[t]
\centering
\includegraphics[width=0.7\figwidth]{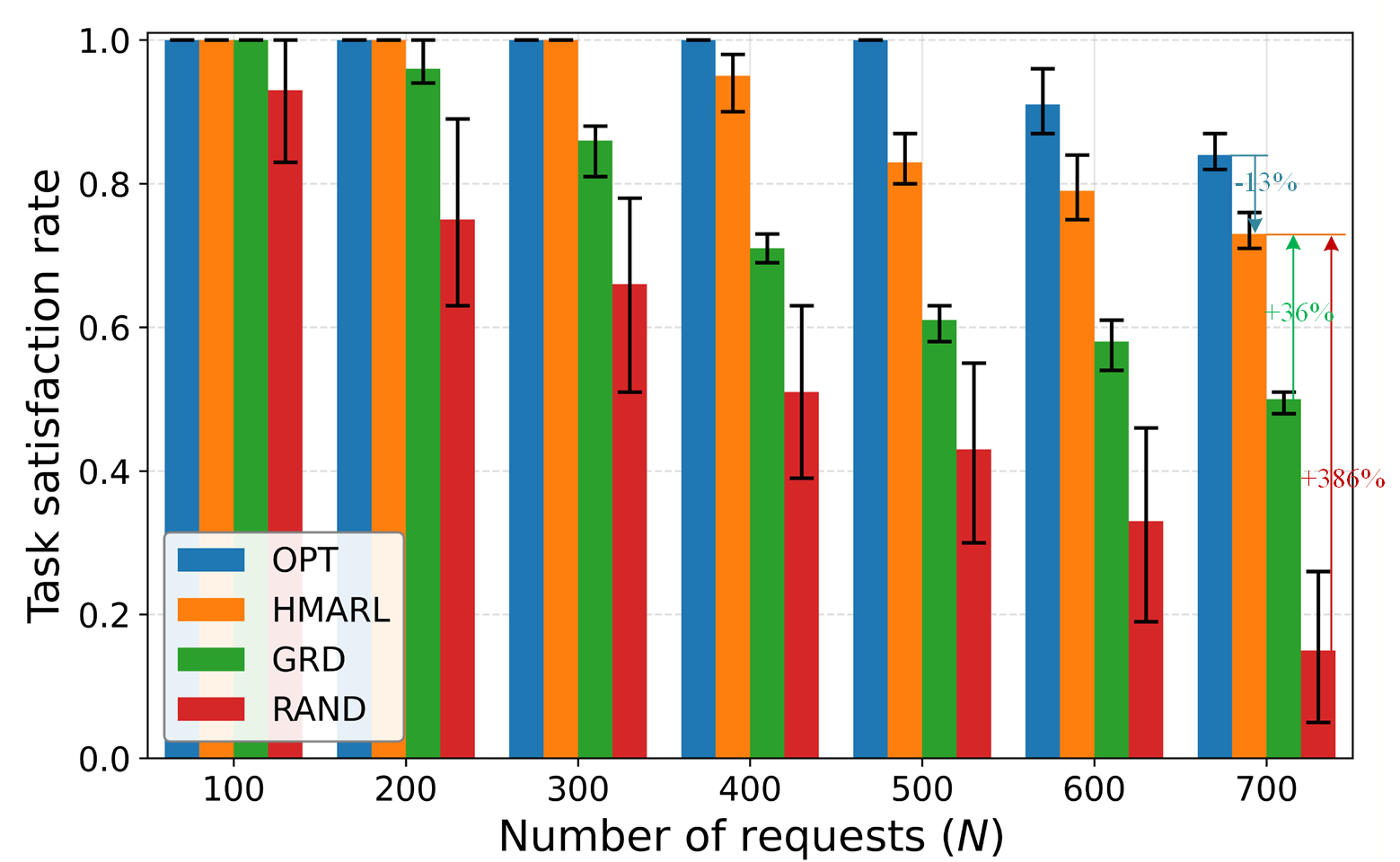}
\vspace{-0.2cm}
\caption{Task satisfaction rate over varying number of requests.}
\label{Fig:4}
\vspace{-0.2cm}
\end{figure}

\begin{figure}[t]
\centering
\includegraphics[width=0.7\figwidth]{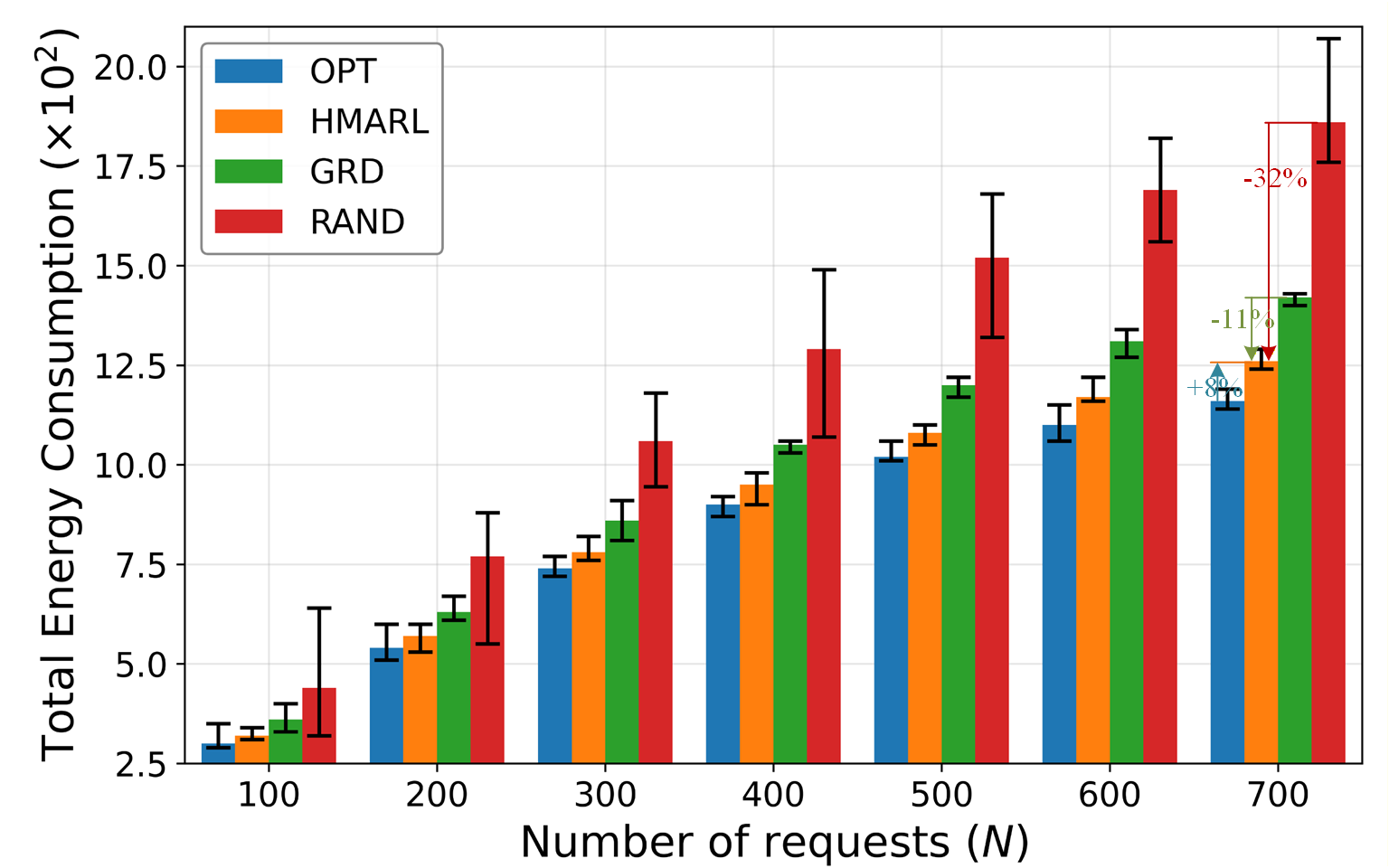}
\vspace{-0.2cm}
\caption{System energy consumption over varying number of requests.}
\label{Fig:5}
\vspace{-0.4cm}
\end{figure}

This section evaluates the HMARL mechanism for dynamic service deployment and traffic distribution.

We consider a system comprising ten terminal-side domains, ten RAN-side domains, and one transport network side domain. The system supports 30 service types. At each time slot, the number of user requests is uniformly sampled from $[N-10,N+10]$, while service types follow a Zipf distribution with an exponent of 1.2. The E2E orchestrator optimizes task distribution by learning a Q-function. Given that service deployment decisions are discrete and traffic distribution decisions are continuous, domain orchestrators employ DDQN agents for service deployment and DDPG agents for traffic distribution. The global objective is to reduce total system energy consumption while improving task satisfaction.

The following baselines are implemented for comparison: (1) RAND: Random strategies for both service deployment and traffic distribution. (2) GRD: A greedy strategy with popularity-based service deployment and load-aware traffic distribution. (3) OPT: An approximate optimal solution obtained via heuristic search using global information. Due to its high computational complexity, OPT is used only for offline benchmarking and is not applicable to online deployment.

Simulation results are shown in Figures \ref{Fig:4} and \ref{Fig:5}. Each result represents the average over 50 consecutive time slots after convergence; bars indicate mean values, and error bars show the range between the maximum and minimum within the window. Overall, HMARL achieves performance close to OPT and consistently outperforms GRD, while RAND exhibits the worst and most unstable performance. Under light load (e.g., $N=100$), all schemes show marginal differences. As the request volume increases, performance gaps become more pronounced. In particular, GRD degrades rapidly due to its reliance on local information. In contrast, HMARL maintains better performance across varying load levels. This advantage stems from joint global–local state awareness and learning-driven cross-domain coordination. By enabling dynamic decision coordination across layers, HMARL adaptively adjusts service deployment and traffic distribution according to request patterns and system status.

We also explored centralized single-agent and fully distributed multi-agent schemes. However, both exhibit instability or convergence issues during training and are therefore excluded from the final comparison, highlighting the importance of hierarchical coordination for scalability and performance.
\section{Challenges and Future Research Directions}
6G CNC orchestrates distributed computing and heterogeneous networks to enable a new service model for telecom operators, efficiently connecting service providers and high-value customers for reliable on-demand E2E services. Despite its potential to transform information infrastructure and drive the digital economy, 6G CNC faces several challenges.

\textbf{Interaction Between Large Models and CNC:}
AI-native intelligence has emerged as a core design paradigm in 6G networks. Recent advances in large foundation models, exemplified by the GPT series, have enabled powerful cross-modal perception and reasoning capabilities, making them promising candidates for the “central brain” of 6G CNC. Within the proposed framework, multi-agent systems (MAS) empowered by large foundation models provide a promising approach to intelligent orchestration. Meanwhile, large-model training and inference impose stringent demands on computing, storage, and network resources, which can be efficiently supported by 6G CNC. This forms a closed-loop interaction: large models drive intelligent decision-making, while CNC provides resource support and environmental feedback for continual optimization. Nevertheless, practical deployment still faces key challenges. Model hallucinations may lead to unreliable orchestration decisions, while the data and compute-intensive nature of large-model workloads requires model-aware orchestration and tight communication-computation co-optimization.

\textbf{Incentive Mechanism Design for Multi-Party Collaboration:}
6G CNC ecosystems involve diverse stakeholders, including users, resource providers, and service operators—each with heterogeneous capabilities, costs, and privacy preferences, resulting in a multi-party, multi-strategy interaction landscape. The coexistence of differentiated objectives necessitates the design of effective incentive mechanisms to promote active cooperation, resource sharing, privacy protection, and system trustworthiness. Such mechanisms must balance individual utility and global efficiency while preventing free-riding and detecting malicious behavior through transparent and verifiable collaboration frameworks. Future research may leverage blockchain and privacy-preserving computation to establish a sustainable and trustworthy CNC ecosystem.

\textbf{Standardization and Implementation Compatibility:}
Although 6G CNC exhibits strong potential for cross-domain coordination at the architectural level, its practical deployment faces significant challenges, particularly in compatibility with existing standards such as O-RAN and cloud-native interfaces. As networks evolve toward openness and softwarization, multiple control and orchestration entities coexist, with notable discrepancies in resource abstraction granularity, interface semantics, and control latency, making it difficult to directly translate E2E orchestration decisions into domain-specific control actions. An approach is to adopt a hierarchical interface mapping mechanism, where high-level orchestration policies are implemented through O-RAN RIC (via rApps/xApps and A1/E2 interfaces) in the RAN and cloud-native orchestration APIs (e.g., Kubernetes) in the computing domain. Future efforts should focus on unified cross-domain resource abstractions, standardized orchestration semantics, and open interface adaptation mechanisms to enhance the deployability, interoperability, and scalability of CNC architectures.

\textbf{Green CNC and Sustainable Development:}
While CNC significantly enhances system capability, it also incurs substantial energy consumption and carbon emissions, especially under compute-intensive workloads such as large models. In 6G, energy efficiency and carbon footprint must be incorporated alongside latency and throughput as key optimization objectives for CNC systems. Future designs should enable carbon-aware resource management by integrating renewable energy sources, leveraging energy-efficient hardware, and coordinating energy optimization across the end–network–cloud continuum. Furthermore, resource scheduling must adapt to dynamic energy supply and demand, promoting the co-evolution of green computing strategies and network orchestration to establish a sustainable, low-carbon foundation for 6G CNC.

\section{Conclusion}
This paper proposed a 6G CNC architecture consisting of three layers: the service orchestration layer, the E2E orchestration layer, and the domain orchestration layer, which together facilitate efficient interaction among service providers, users, and infrastructure providers. Building on this architecture, a hierarchical multi-agent-based orchestration mechanism was developed to enable cross-domain resource coordination. Extensive simulation results demonstrate that the proposed approach consistently outperforms baseline solutions in terms of task satisfaction and system energy efficiency. Finally, this paper discussed key challenges and future research directions for 6G CNC, providing valuable insights into the design and future evolution of CNC systems.


\bibliographystyle{IEEEtran}
\bibliography{main}

\section*{Acknowledgment}

This work was supported by the National Science Foundation of China under Grant 62271062.

\section*{Biography}
\setlength{\parskip}{0pt} 
\setlength{\parindent}{0pt} 

\newcommand{\biobreak}{\vspace{6pt}} 


\textbf{Yang Li} (ly209991@bupt.edu.cn) is currently pursuing the Ph.D. degree at Beijing University of Posts and Telecommunications (BUPT), China. His research interests include mobile edge networks and computing power networks.

\biobreak

\textbf{Xing Zhang} (zhangx@ieee.org) is a full professor with the School of Information and Communications Engineering, BUPT, China. His research interests are mainly in 5G/6G networks, satellite communications, edge intelligence, and Internet of Things. He is a senior member of IEEE.

\biobreak

\textbf{Yan Zhang}  (yanzhang@ieee.org) is a full professor with the School of Information and Communication Engineering, University of Electronic Science and Technology of China. His research interests include next-generation wireless networks leading to 6G, and green and secure cyber–physical systems. He is a Fellow of IEEE and IET.

\biobreak

\textbf{Wenbo Wang}  (wbwang@bupt.edu.cn) is a former vice president and a professor at BUPT, a Fellow of the China Institute of Communications. His current research interests include radio transmission technology, wireless network theory, cognitive communications, and software radio technology. He is a senior member of IEEE.

\end{document}